# High-performance source of spectrally pure, polarization entangled photon pairs based on hybrid integrated-bulk optics

EVAN MEYER-SCOTT,* NIDHIN PRASANNAN, CHRISTOF EIGNER, VIKTOR QUIRING, JOHN M. DONOHUE, SONJA BARKHOFEN, AND CHRISTINE SILBERHORN

*Integrated Quantum Optics, Department of Physics, University of Paderborn, Warburger Straße 100, 33098 Paderborn, Germany*

*evan.meyer.scott@upb.de

**Abstract:** Entangled photon pair sources based on bulk optics are approaching optimal design and implementation, with high state fidelities, spectral purities and heralding efficiencies, but generally low brightness. Integrated entanglement sources, while providing higher brightness and low-power operation, often sacrifice performance in output state quality and coupling efficiency. Here we present a polarization-entangled pair source based on a hybrid approach of waveguiding and bulk optics, addressing every metric simultaneously. We show 96 % fidelity to the singlet state, 82 % Hong-Ou-Mandel interference visibility, 43 % average Klyshko efficiency, and a high brightness of $2.9 \times 10^6$ pairs/(mode·s·mW), while requiring only microwatts of pump power.

## 1. Introduction

Modern entangled photon pair sources based on parametric down-conversion (PDC) are approaching the ideal: high state fidelity, spectral purity, and heralding efficiency are commonly demonstrated, enabling applications such as tests of Bell's inequality [1,2], probing the boundaries of quantum physics [3, 4], quantum communication over long distance [5–7], and quantum metrology beyond classical limits [8–10]. However, sources using bulk nonlinear crystals suffer an intrinsic three-way tradeoff between brightness, fiber-coupling efficiency, and spectral purity [11]. This deficiency is now becoming critical, as many new experiments and applications rely on the interference of multiple photons. For high rate, high quality multi-photon experiments, all three of the aforementioned parameters must be maximized simultaneously [12–18]. This is because the overall rate in multi-photon experiments with $N$ pairs scales with probability $p$ of producing and detecting a single pair as $p^N$, requiring both high brightness and coupling efficiency. The quality of multi-photon interference is determined by spectral purity and indistinguishably as only pure, indistinguishable photons are able to interfere with high visibility.

In contrast to bulk sources, integrated sources provide high brightness due to strong confinement in waveguides and long interaction lengths, and can be designed to be spatially and spectrally single-mode, enabling simultaneously high fiber-coupling efficiency, spectral purity, and brightness. Many examples exist of high-brightness integrated sources, for example based on PDC in waveguides [19–21], or four-wave mixing in optical fibers [22,23] and silicon waveguides [24–27]. However, these sources have not yet demonstrated simultaneous high performance in all other parameters comparable to their bulk-optical counterparts. Entangled pairs from quantum dots, while promising [28, 29], also do not yet reach the performance of pairs from nonlinear optical sources.

Here we solve the performance problem of integrated optics while retaining the coupling and brightness benefits by employing a hybrid bulk-waveguide solution: photon pairs are produced in a single-mode waveguide, then made to interfere and coupled to optical fiber using bulk optics.

Our source combines for the first time high performance in all parameters simultaneously.

## 2.    Integrated single-mode photon pair sources

Over the last two decades, efforts in improving entangled photon-pair sources based on bulk crystals and bulk optics have resulted in impressive performance in many measures (see Table 1 in the appendix for comparison). Entanglement fidelities above 99 % are readily achieved [1,2,17,30], and even above 99.9 % is possible [4]. Klyshko (heralding) efficiency [31], defined as the ratio of coincidence to singles counts, can reach 75 %  [1, 2, 30, 32, 33]. The spectral purity, required to interfere photons from separate sources for multi-photon experiments, has been shown above 99 % [34].

Unfortunately, bulk sources suffer from an intrinsic tradeoff between the brightness, or emitted photon rate per pump power (taken in the source before losses, but considering only modes which will reach the detectors), and the Klyshko efficiency [11]; for example setting the pump focus to enable coupling photon pairs to single mode fiber with 95 % efficiency necessarily reduces the brightness by a factor of ten from the maximum [35]. This tradeoff arises due to conflicting requirements on the focusing conditions: high brightness requires a tight pump focus which concentrates the down-converted light into the spatial modes collected by the fibers [36]. High Klyshko efficiency, however, requires a weak focus which more strongly correlates the spatial modes of signal and idler photons such that if one photon is coupled into fiber, the other is likely to be coupled too [37]. This tradeoff means the fundamental performance limits of bulk sources have largely been saturated. Furthermore, sources at telecommunications wavelengths are much less bright than those with visible-range photons, due to the wavelength dependence of the down-conversion efficiency [38].

Integrated photon sources can surpass these limits, as the waveguide, rather than spatial phasematching, defines the allowed modes into which photons are emitted [39]. Single-spatial-mode waveguides in particular completely decouple the brightness from the focusing conditions [40], and can be produced with appropriate choice of the waveguide width and height. Then the the maximum coupling efficiency depends only on the mode overlap between the waveguide and fiber modes. While the brightness of bulk sources with optimal focusing scales with increasing nonlinear crystal length $L$ as $\sqrt{L}$ [41] or constant [11], the brightness of waveguide sources increases proportionally to $L$, as well as inversely with the effective area [38]. This allows waveguide sources, beyond removing the brightness-efficiency tradeoff, to be significantly brighter overall [42], requiring much less pump power for reasonable photon pair probabilities (e.g. 2 mW average power for 0.1 pairs generated per pulse using waveguides [43] vs 660 mW in bulk [17]).

It is also desirable to have spectrally single-mode photons, where a frequency measurement of the signal provides no information on the properties of the idler, meaning each is in a pure spectral state. This is required for interference between independent sources, essential for quantum networking [15, 16, 44, 45], boson sampling [12, 13] or linear optic quantum computing [12–14]. This high spectral purity can be asymptotically accomplished by narrowband filtering, but filtering both photons unavoidably lowers the Klyshko efficiency [46,47]. Engineering the group velocities of the pump, signal, and idler avoids this problem by producing intrinsically single-spectral-mode photons [48, 49]. However, for bulk crystals, even a spectrally-engineered source has only a certain range of focusing parameters where the spectral purity is maximized [11, 37, 50]. In waveguides, this spectral-spatial coupling is eliminated thanks to the single-spatial-mode propagation, allowing the spectral purity to be independently optimized.

Yet to date the most advanced experiments do not use waveguide sources, which can be understood in light of the difficulty in optimizing performance in integrated optics (see Table 2 in the appendix for comparison). The brightness of integrated sources is orders of magnitude larger than possible in bulk, reaching above $10^8$ pairs/(s·mW) [21,23]. Entanglement fidelity over 95 % has been achieved in a few integrated systems [20, 23, 51, 52], and while source engineering has allowed high spectral purity in waveguided photon pair sources [40, 53, 54], so far it has not been combined with qubit entanglement, with the exception of temporally filtered systems [55]. The biggest drawback of current implementations however is the Klyshko efficiency, which is often very low due to lossy integrated components and poor coupling between elements. Though high Klyshko efficiencies from unentangled waveguide sources have been demonstrated [43], only one example of an entangled pair source with efficiency >5 % exists so far [56].

By combining both integrated and bulk approaches we benefit from the advantages of waveguide photon pair sources – single mode operation, high brightness, independent optimization of parameters – and the flexibility of efficient free-space coupling to fiber. We describe the experimental setup and results below.

## 3. Experiment

Our hybrid source of entangled photon pairs is based on a free-space-coupled waveguide in a periodically poled potassium titanyl phosphate (KTP) crystal. In fact we test two such chips (labeled 1 and 2), and find that chip 1 has better coupling efficiency and chip 2 better entanglement properties. We present all data and figures (except Fig. 2) for chip 1, and discuss the differences between the samples in Section 3.4. The periodic poling is designed for type-II phasematching at 1550 nm, and the material KTP is chosen such that the group velocity of the pump is between that of the signal and idler. Matching the pump and phasematching bandwidths then provides intrinsic spectral purity of the photons [40, 57].

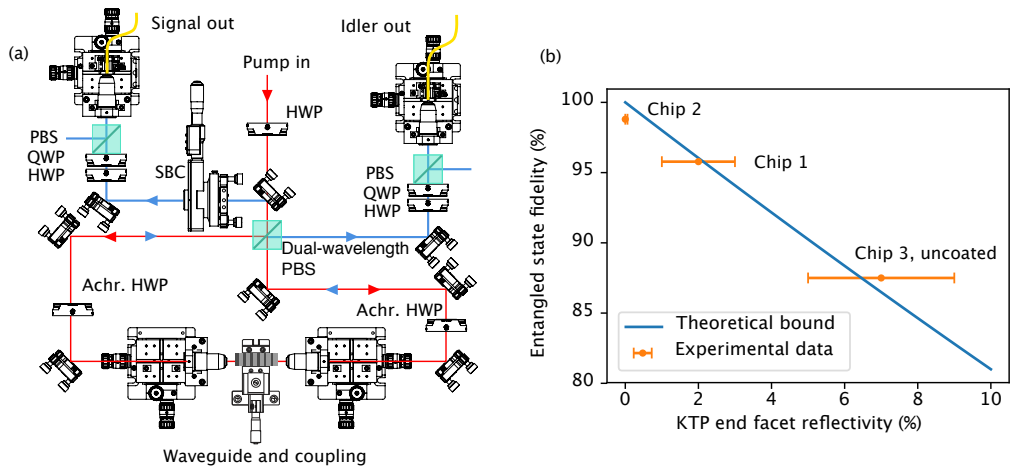

Fig. 1. (a) Experimental setup of hybrid integrated-free-space source. We present the important optical and mechanical elements in the actual layout, to scale, such that the setup can be easily reproduced. HWP - half-wave plate, PBS - polarization beam splitter, SBC - Soleil-Babinet compensator. (b) Maximal theoretical bound of achievable entangled state fidelity given nonzero end facet reflectivity in the KTP waveguide, and experimental data points for different chips.

### 3.1. Setup

The KTP waveguide is placed in a Sagnac loop [58, 59] as shown in Fig. 1(a). The pump (Coherent Mira 900f) at 770 nm is coupled through a single-mode fiber for spatial mode cleaning, then its polarization is set with a half-wave plate (HWP), and it passes through a dichroic mirror (Thorlabs DMSP1180). The pump is split equally at a dual-wavelength polarization beam splitter (PBS, OptoSigma), and propagates both ways around the Sagnac loop. In the counter-clockwise direction the pump polarization is rotated to horizontal with a superachromatic HWP (B. Halle), balanced with an identical HWP at 0° on the clockwise path to reduce distinguishably from dispersion. Then the pump is coupled using achromatic lenses of focal length 6 mm (Edmund Optics ACH-NIR 6 X 9 NIR-II) into the KTP waveguide (ADVR, Inc.) with length 9 mm. For producing a single spectral mode, this waveguide length requires a pump bandwidth of 1.8 nm, which we set using a 4f-line as a variable bandpass filter in the pump path. The same lenses couple out the photon pairs, where now the clockwise-propagating pairs are rotated 90° in polarization by the HWP. These interfere with the un-rotated counter-clockwise pairs at the dual-wavelength PBS, which divides signal and idler to the two output ports while creating the polarization entangled-state $|\psi\rangle = \frac{1}{\sqrt{2}}\left(|HV\rangle + e^{i\phi}|VH\rangle\right)$. The phase $\phi$ of the state is set to $\pi$ in the signal arm using a Soleil-Babinet compensator (Thorlabs, SBC-IR), and both photons pass filters to remove the pump (anti-reflection coated silicon), and the sinc lobes of the PDC spectrum and fluorescence (signal: Semrock NIR01-1550/3, idler: Thorlabs FBH1550-12, chosen due to the slight asymmetry in photon bandwidth). Finally the entangled state is analyzed with half- and quarter-wave plates and PBSs, then coupled into optical fiber for detection with superconducting nanowire single photon detectors (Photon Spot, Inc.).

### 3.2. Differences to bulk sources

The single-mode nature of the PDC in our waveguide brings significant advantages over bulk optics. Since the clockwise and counterclockwise photons are necessarily in the same spatial mode, interference at the PBS is simple, and alignment is straightforward. This also relaxes much of the strict symmetry needed in the crystal position in bulk Sagnac sources with respect to the focusing lenses and PBS [60]. The high brightness of the source allows for low pump power, which means the pump spatial mode can be cleaned in standard single mode fiber without spectral broadening due to nonlinearities. One drawback is the lenses used to couple light in and out of the waveguide are the same for pump and photon pairs, meaning the focus is optimized only for the down-converted photons. Nonetheless, with achromatic lenses we can reach >40 % coupling of the pump through the waveguide.

There is one point in waveguides that requires special attention compared to bulk sources: the anti-reflection coating on the crystal surface. In bulk sources, photons reflected internally at the end facets have a different focal position when they reach the coupling fibers, and thus couple poorly. In waveguides by contrast, photons reflected at the end facets remain in the single spatial mode and couple well to the fibers. Unfortunately these photons end up in exactly the wrong polarization compared to their non-reflected partners, directly lowering the entanglement fidelity as in Fig. 1(b). We solve this using an ion-assisted coating technique to deposit anti-reflection coatings for both wavelengths on both end facets. These coatings reduce the end facet reflectivity over a bandwidth of 100 nm to around 2 % for the chip 1 and below the measurement uncertainty of 0.06 % for the chip 2, giving a maximum achievable fidelities of 96 % and 99.9 % respectively.

### 3.3. Distinguishability in time and frequency

For any polarization-entangled photon pair source, it is essential that the two polarization paths are completely indistinguishable in all other degrees of freedom. The single-mode waveguide and output fiber coupling ensure this indistinguishability in the spatial degree of freedom, but extra care must be taken to ensure time-frequency overlap, particularly as spectrally pure photons require relatively broadband pump pulses, especially compared to continuous-wave sources. The Sagnac scheme does not require degenerate signal and idler emission, but does require that the clockwise (c) and counter-clockwise (cc) paths remain indistinguishable. Even though both paths encounter the exact same optical components, they encounter them at different wavelengths and polarizations (e.g. pump vs photon wavelength, signal vs idler polarization). Any uncompensated dispersive or birefringent materials or coatings thereby reduce the polarization entanglement generated by coupling polarization information to the time-frequency degree of freedom.

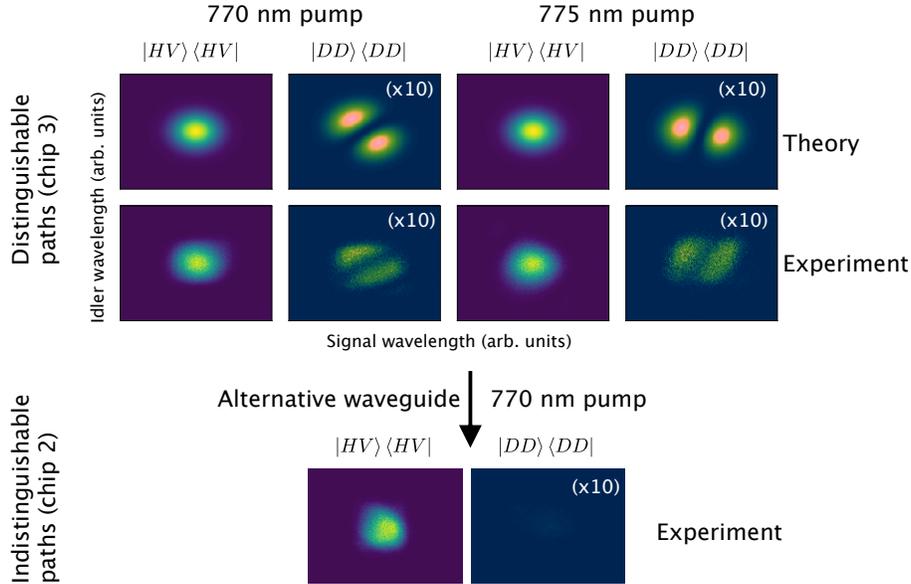

Fig. 2. Investigating the distinguishability of photon pairs via polarization-resolved joint spectra. We show the joint spectral intensity as simulated (top) and measured (bottom) when projecting onto the polarization state $|HV\rangle$ or $|DD\rangle$. The $|DD\rangle$ plots have been scaled up in intensity by a factor 10 for readability. These diagonal measurements exhibit fringing owing to time-frequency distinguishability between the two paths of the Sagnac source, resulting in reduced overall visibility since the detectors are not sensitive to this time-frequency information. By changing the wavelength of the PDC pump (left vs. right), we see a change in the orientation of the fringes, suggesting the presence of a direction-dependent chirp on the pump. To eliminate the distinguishability and therefore the undesired counts in the D/A basis (plot shown to scale with the other D/A measurements) we found different waveguides in chips 1 and 2 with more uniform phasematching.

This coupling can be modelled in the joint polarization and time-frequency space for signal and idler photons in modes defined by creation operators $\hat{a}$ and $\hat{b}$, respectively, as

$$|\psi\rangle = \frac{1}{\sqrt{2}} \int d\omega_s d\omega_i \Big( f_{cc}(\omega_s, \omega_i) \hat{a}^\dagger_{H,\omega_s} \hat{b}^\dagger_{V,\omega_i}$$
$$- f_c(\omega_s, \omega_i) \hat{a}^\dagger_{V,\omega_s} \hat{b}^\dagger_{H,\omega_i} \Big) |00\rangle . \tag{1}$$

This model state always has perfect anti-correlations in the rectilinear (H/V) basis. The projection probability in the diagonal basis is

$$\left| \left( \frac{\langle H| + \langle V|}{\sqrt{2}} \right) \left( \frac{\langle H| \pm \langle V|}{\sqrt{2}} \right) |\psi\rangle \right|^2$$
$$= \frac{1}{8} \int d\omega_s d\omega_i |f_{cc}(\omega_s, \omega_i) \mp f_c(\omega_s, \omega_i)|^2 . \tag{2}$$

If the joint spectral amplitudes for the clockwise and counter-clockwise paths $f_c(\omega_s, \omega_i)$ and $f_{cc}(\omega_s, \omega_i)$ are exactly identical, the polarization and time-frequency degrees of freedom are separable, and we obtain a "perfect" polarization entangled state, with anti-correlations in the diagonal (D/A) basis. For distinguishable paths with $f_c(\omega_s, \omega_i) \neq f_{cc}(\omega_s, \omega_i)$, undesired coincidence counts will be measured when projecting on $|DD\rangle$, as described by Eq. (2). Notably, this projection is sensitive to spectral phase differences between the two paths.

To diagnose sources of distinguishability, we spectrally resolve the polarization correlations [61, 62]. In Fig. 2, we show the joint spectral intensities reconstructed using time-of-flight spectrometers [63, 64] when projecting on polarization combinations in the rectilinear and diagonal polarization bases. The results can be explained by relative time delays $\tau_\ell$ ($\ell = p, s, i$) between the two directions for the signal, idler, and pump, defined as

$$f_{cc}(\omega_s, \omega_i) = e^{i\tau_i \omega_i + i\tau_s \omega_s + i\tau_p(\omega_s + \omega_i)} f_c(\omega_s, \omega_i), \tag{3}$$

which could arise due to spectral phase profiles specific to the vertical ports or facet coatings of the dual-wavelength PBS, or due to unpoled regions on one end of the waveguide. By projecting onto the diagonal basis, these time delays will manifest as fringing across the signal-idler joint spectral intensity oriented at an angle $\theta = \arctan\left(\frac{\tau_p + \tau_i}{\tau_p + \tau_s}\right)$. If a relative chirp exists between the two paths, this will appear as a frequency-dependent time delay, $\tau_\ell = \tau_0 + 2A_\ell\delta\omega_\ell$, and the angle of the fringes will change as the central frequencies are shifted, which we observe when the pump wavelength is shifted in Fig. 2. The theoretical simulations in Fig. 2 correspond to a relative time delay of 600 fs between horizontal and vertical components ($\tau_s = -\tau_i = 300$ fs), equivalent to the birefringent delay of approximately 2.2 mm of KTP. To describe the dependence on pump wavelength, a chirp on the pump for the counter-clockwise process of $A_p = 5800$ fs$^2$ is sufficient. To optimize the indistinguishably, we use this polarization-resolved joint spectral characterization to identify waveguides with more uniform poling, and obtain the results presented below.

### 3.4. Results

We measure the spectral purity of our source via the joint spectral intensity (JSI) and Hong-Ou-Mandel (HOM) interference [65] between independent photons. The joint spectral intensity of Fig. 3 is reconstructed using a time-of-flight spectrometer [63, 64], and returns an upper bound to the purity of 98 %.

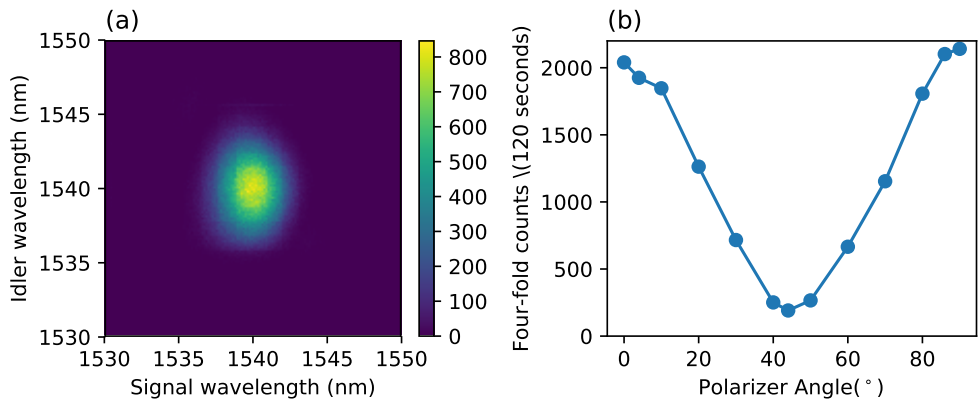

Fig. 3. (a) Joint spectral intensity measured from our KTP waveguide (chip 1) with 770 nm pump, including 8 nm and 12 nm FWHM spectral filters for the signal and idler respectively, giving an upper bound of the spectral purity of 98 %. (b) Hong-Ou-Mandel interference of two independently heralded photons from our waveguide, with visibility (82 ± 2) %.

To measure HOM interference we use the Sagnac loop to create two photon pairs without polarization entanglement, one in each direction around the loop. This is accomplished by detecting two signal photons simultaneously, one $|V\rangle_s$ and one $|H\rangle_s$, making use of both output ports of the signal photon's PBS. This heralds two idler photons $|1H, 1V\rangle_i$ in the same spatial mode heading toward the idler's PBS. To make these photons interfere we rotate the idler HWP, which at 22.5° leads to the state (in a single spatial mode) $\frac{1}{\sqrt{2}} (|2H\rangle_i - |2V\rangle_i)$ due to the indistinguishability of the two heralded photons [66]. Thus we should be able to tune the coincidence probability between the horizontal and vertical output ports of the idler's PBS between 0 and 1 by rotating the HWP. This is in contrast to typical HOM interference, where distinguishability in the photons is introduced via a time delay, and they impinge from separate ports on a 50:50 beam splitter. In that case the coincidence probability varies between 0 and 1/2. Changing the HWP in our case is like changing the splitting ratio of the beam splitter from 100:0 to 50:50 to 0:100. Mapping the HOM visibility from the temporal to the polarization case gives $V_{HOM,pol} = \frac{N_{max}/2 - N_{min}}{N_{max}/2}$, where $N_{max}$ and $N_{min}$ are the maximum and minimum number of fourfold coincidences we measure, respectively, and the factor one half comes from the maximum probability being 1 compared to 1/2 in the temporal case (see appendix). This visibility depends on the spectral purity and indistinguishably of the photons [67], and can also be degraded by higher-order down-conversion events. Our measured value is $V_{HOM,pol} = (82 \pm 2)\,\%$, without background subtraction, for 0.003 pairs per pulse. That the HOM visibility is lower than the upper bound given by the JSI's purity can be fully explained by backreflections from the waveguide's end facets as in Fig. 1(b), to which the HOM interference is much more sensitive than the polarization entanglement. A simulation [68] with 2 % reflection on each end facet into a temporally orthogonal mode (due to the extra traversal of the crystal for reflected photons) returns a predicted HOM visibility 81.9 %, which agrees with the experiment, and implies other sources of imperfection are negligible.

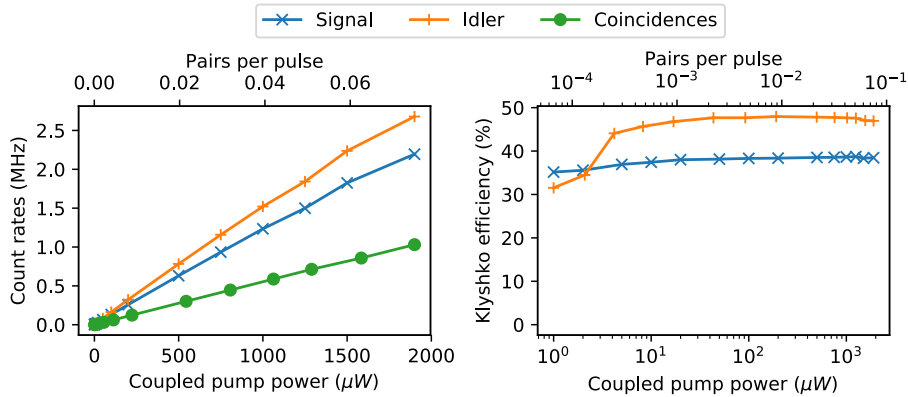

Fig. 4. (left) Single and coincidence count rates vs pump power exiting the waveguide (chip 1), from which we extract a brightness of $(3.5 \pm 0.2) \times 10^6$ pairs/(s·mW). (right) Klyshko efficiencies vs pump power, calculated from coincidences divided by singles. The average efficiencies (excluding the first two and last five points) are $(38.0 \pm 0.5)\,\%$ and $(46.8 \pm 1.3)\,\%$ for the signal and idler respectively. Uncertainties in both cases are the standard deviations of the multiple data points.

We plot the measured count rates and Klyshko efficiencies in Fig. 4. The rates scale linearly with pump power (at low power), allowing us to plot also versus the mean pair number per pulse produced in the crystal. We are limited to around 2 mW pump power and 1 million coincidences per second by saturation and latching of our detectors. From Fig. 4(a) we extract a brightness of $(3.5 \pm 0.1) \times 10^6$ pairs/(s·mW), competitive with state-of-the-art waveguide processes. The Klyshko efficiencies, as expected, drop at low power due to dark counts, and at high power due to detector saturation. The average Klyshko efficiencies (excluding the first two and last five points) are $(38.0 \pm 0.5)\,\%$ and $(46.8 \pm 1.3)\,\%$ respectively for signal and idler. We can compare these values to those estimated from classical measurements: component transmission from waveguide to fiber for the signal $(85 \pm 3)\,\%$ (idler $(79 \pm 3)\,\%$), fiber coupling efficiency for the signal $(66 \pm 5)\,\%$ (idler $(84 \pm 5)\,\%$), fiber transmission to detectors $(95 \pm 5)\,\%$, detector efficiency $(90 \pm 5)\,\%$, giving a total expected efficiency of $(48 \pm 4)\,\%$ for the signal and $(57 \pm 6)\,\%$ for the idler. Waveguide losses partially account for the difference to the measured efficiencies, as do losses due to the gentle filtering used to remove sinc lobes [47]. None of these losses are fundamental: better coatings on our optics and lower loss waveguides would boost the efficiency dramatically. Additionally, the heralded $g_h^{(2)}(0)$, an indication of noise photons in the system, is consistent with zero extra noise. For spectrally pure photons and low pump power, $g_h^{(2)}(0) = 2(2 - \eta_h)\mu$, where $\mu$ is the mean pair number per pulse and $\eta_h$ is the Klyshko efficiency of the heralding signal photon. This gives $g_h^{(2)}(0)/\mu = 3.24$, which agrees with our experimental result of $3.19 \pm 0.05$.

Finally we present the entanglement visibility curves and reconstructed two-qubit density matrix in Fig. 5. We find a maximum visibility of $(96.0 \pm 0.1)\,\%$ in the rectilinear basis and $(94.3 \pm 0.1)\,\%$ in the diagonal basis, where the error bars come from Poissonian statistics. We then perform overcomplete quantum state tomography [69] with an average coincidence rate of $59\,000$ pairs/s, finding a fidelity of $F = \langle\psi^-|\,\rho\,|\psi^-\rangle = (95.78 \pm 0.04)\,\%$, and tangle $0.842 \pm 0.100$. These are limited by the small end facet reflectivity as in Fig. 3(b), some residual temporal-spectral distinguishability as in Section 3.3, and 30 accidental coincidences per second due to multi-pair emissions.

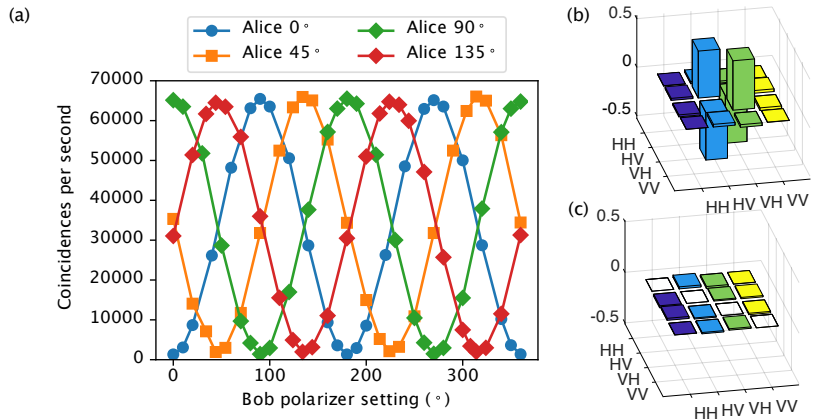

Fig. 5. (a) Correlation curves of coincidence counts from chip 1. (b) Real, and (c) imaginary parts of the reconstructed density matrix, giving $(95.78 \pm 0.04)\,\%$ fidelity to $|\psi^-\rangle$.

We also measured a different waveguide chip (chip 2) with the same design, but from a different batch. Due to better end coatings, the entanglement fidelity of chip 2 was $(98.82 \pm 0.05)\,\%$, the tangle $0.9841 \pm 0.0006$, and the Hong-Ou-Mandel visibility $(89 \pm 2)\,\%$. Unfortunately the waveguide and coupling losses were higher, resulting in Klyshko efficiencies of $(26.3 \pm 0.3)\,\%$ and $(28.2 \pm 0.5)\,\%$. Thus we chose to focus on chip 1, with the best Klyshko efficiency, at the cost in this case of entanglement and Hong-Ou-Mandel visibility. However both the reduced visibility of the first chip and the reduced efficiency of the second chip are technical, rather than fundamental problems, which will be improved in our case by bringing fabrication in house, where we can better control the process.

### 3.5. Comparison with other sources

It is instructive to compare the properties and performance of our source to previous work. Compared to bulk sources (see Table 1 in the appendix), our waveguide source provides $1 - 3$ orders of magnitude higher brightness, as well as higher Klyshko efficiency than the bright bulk sources due to the waveguide decoupling the brightness and efficiency. Our entanglement fidelity and HOM interference visibility are comparable to many of the best bulk sources.

Compared to integrated sources (see Table 2 in the appendix), our source has by far the highest Klyshko efficiency, in some cases by nearly two orders of magnitude. Our high spectral purity leads to an interesting brightness comparison: instead of the typical pairs/(s·mW), we can compare the brightness additionally per spectral mode. Our brightness by this metric is still very high ($2.9 \times 10^6$ pairs/(mode·s·mW)), while CW pumped sources and those without spectral engineering drop by orders of magnitude. Our source is the brightest to emit entangled pairs into a single spectral-temporal mode, suitable for multi-photon interference.

## 4. Conclusion

We have demonstrated a source of entangled photon pairs that seeks to simultaneously maximize polarization entanglement, spectral purity, and brightness, and have shown for the first time high Klyshko efficiency in a waveguided entangled pair source. Further optimization of optical coatings and beam reshaping to maximize the overlap between waveguide and fiber modes [70] will allow coupling efficiencies approaching 100 %, independent of the source brightness. The entanglement and HOM interference visibilities we measured indicate a good anti-reflection coating on the waveguide end facets is critical, something we will improve in future work. Our use of spectral engineering to produce pure, indistinguishable photons is a great advantage over spectral filtering for quantum networking, and could be further enhanced by apodization of the nonlinearity as shown in bulk sources [71–73]. We are also examining the fabrication processes responsible for variability in waveguides, seeking to find operational points that minimize the effect of waveguide and poling fluctuation [74]. Overall we have taken a significant step towards the ideal integrated source of entangled photons, and laid out the specific optimizations necessary for maximum performance.

Looking forward, the high brightness, fidelity and purity of our source make it an excellent candidate for multiplexing to create multi-photon entangled states [17, 18], and for time-multiplexed multi-photon experiments [75]. The wavelength is also compatible with telecom infrastructure, making the source suitable for quantum communications in optical fiber, like teleportation and entanglement swapping. Another exciting application is polarization squeezing [76], which requires simultaneously high brightness and coupling efficiency, to produce squeezed and entangled continuous-variable states that can be detected without a local oscillator [77, 78].

## Appendix

### HOM interference visibility for polarization

The standard HOM interference visibility between two individual photons incident on seprate ports of a 50:50 beamsplitter is defined as $V_{HOM} = \frac{I_{max} - I_{min}}{I_{max}}$, where $I_{max}$ is the probability of getting a coincidence when the two photons are made distinguishable by e.g. a time delay, and $I_{min}$ is the coincidence probability when the distinguishability is removed, e.g. by maximal temporal overlap of the photons. For two single photons given by states $\rho_1$ and $\rho_2$, $I_{max} = 0.5$ as each is separately sent to each detector with 50 % probability, and $I_{min} = \frac{1}{2} (1 - \text{Tr}[\rho_1 \rho_2])$, given by the purity and indistinguishability of the photons [67], resulting in $V_{HOM} = \text{Tr}[\rho_1 \rho_2]$.

If the same photons are prepared instead in a single spatial mode in orthogonal polarizations, then made to interfere with a HWP and PBS, the probabilities are somewhat different. In particular, now $I_{max,pol} = 1$ due to the perfect splitting of the horizontal and vertical photons at the PBS. However, we still find $I_{min,pol} = I_{min}$ due to the equivalence of the 50:50 beam splitter and HWP at 22.5°. The beam splitter transformation for 50:50 splitting is $\frac{1}{\sqrt{2}} \left( \begin{smallmatrix} 1 & 1 \\ -1 & 1 \end{smallmatrix} \right)$, which equals the rotation matrix for the HWP at 22.5°. To ensure that $V_{HOM} = V_{HOM,pol}$ for the same photons, we thus must take $V_{HOM,pol} = \frac{I_{max,pol}/2 - I_{min,pol}}{I_{max,pol}/2}$. Taking these probabilities to numbers of coincidence counts gives us the polarization HOM visibility in the main text.

*Comparison tables*

Table 1 contains a comparison of high-performance entangled-pair sources based on bulk optics. In both tables, † denotes photons at telecommunications wavelengths; * denotes values inferred from published data; MZ stands for Mach-Zehnder; and WG stands for waveguide. In particular, entanglement fidelity is approximated from reported visibilities as $F = 1 - (1 - V)/2$, where $V$ is the highest reported average visibility in rectilinear and diagonal bases without subtracting accidental coincidences, and HOM visibility between two individually heralded photons is upper bounded by the spectral purity. The brightness encompasses the filtered source bandwidth, and all losses are removed to give the brightness inside the crystal. The Klyshko efficiency is the average of the signal and idler photons, evaluated as the coincidence rate divided by the singles rate. Values that could not be estimated are indicated with –.

Table 2 contains a comparison of high-performance entangled pair sources based on integrated optics, with the same conventions as above. For Jöns17, the Klyshko efficiency is replaced by the approximate extraction times detection efficiency, while for Huber17 it is upper-bounded by the extraction efficiency. The HOM visibility for Huber17 is given by the average two-photon interference visibility for QD2.

Table 1. **Bulk optics, see description above.**

| Reference | Architecture | Entanglement fidelity | HOM visibility | Brightness (pairs/(s·mW)) | Klyshko efficiency |
|---|---|---|---|---|---|
| **CW-pumped** | | | | | |
| Kwiat95 [79] | BBO, rings | 98.9%* | – | 1000* | 10%* |
| Kwiat99 [80] | Crossed BBO | 99.4%* | - | 2400* | 6.5%* |
| Fiorentino04 [81] | PPKTP MZ | 95%* | - | 370 000 | 18% |
| Kim06 [59] | PPKTP Sagnac | 99.1%* | - | 200 000* | 16%* |
| Steinlechner14 [82] | PPKTP Sagnac | 99.4% | - | 47 000 | 38% |
| Poh15 [4] | BBO, rings | 99.99%* | - | 9000* | 6.3%* |
| Chen18 [83] | Crossed PPKTP + Sagnac | 99.2% | - | $4.7 \times 10^6$ | 18.5% |
| **Pulsed-pumped** | | | | | |
| Christensen13 [30] | Crossed BiBO | 99.8%* | – | – | 75% |
| Giustina15 [1] | PPKTP Sagnac | 99.5%* | – | – | 77% |
| Shalm15 [2] | PPKTP small MZ[†] | 99.9%* | – | – | 75% |
| Wang16 [17] | Crossed BBO | 99.5% | 91% | 12 000 | 42%* |
| **Spectrally engineered** | | | | | |
| Evans10 [84] | PPKTP small MZ[†] | 95%* | 93%* | 123 000 | 1.9% |
| Jin14 [85] | PPKTP Sagnac[†] | 97.3% | 82%* (theory) | 200 000* | 10% |
| Weston16 [34] | PPKTP Sagnac[†] | 99.0% | 100% | 1500* | 52% |
| **This work Chip 1** | PPKTP WG + Sagnac[†] | 95.8% | 82% | $3.5 \times 10^6$ | 43% |
| **Chip 2** | PPKTP WG + Sagnac[†] | 98.8% | 89% | $5.6 \times 10^6$ | 27% |

Table 2. **Integrated optics, see description above.**

| Reference | Architecture | Entanglement fidelity | HOM visibility | Brightness (pairs/(s·mW)) | Klyshko efficiency |
|---|---|---|---|---|---|
| **CW-pumped** | | | | | |
| Herrmann13 [19] | PPLN, double-poling[†] | 97.5%* | - | $4 \times 10^5$ | 2%* |
| Clausen14 [52] | Free space MZ | 97.9%* | 91%* | $6.6 \times 10^5$ | 5% |
| Autebert16 [86] | AlGaAs Bragg[†] | 93.4%* | - | – | 1%* |
| Vergyris17 [21] | PPLN, fiber Sagnac[†] | 92.6%* | - | $2.4 \times 10^8$ | 3%* |
| Atzeni18 [87] | 2x PPLN WG + BS[†] | 92.9% | - | $2.2 \times 10^9$* | 0.04% |
| **Pulsed-pumped** | | | | | |
| Li05 [22] | Fiber Sagnac[†] | 65%* | – | $8 \times 10^7$* | 0.6%* |
| Fan07 [23] | Microstructure fiber Sagnac | 96.3% | – | $8 \times 10^8$* | 0.6% |
| Lim08 [88] | PPLN, fiber Sagnac[†] | 96.8% | – | – | 0.6%* |
| Arahira11 [51] | PPLN, fiber Sagnac[†] | 99.6%* | – | $5 \times 10^6$* | 0.6%* |
| Sansoni17 [20] | PPLN MZ[†] | 97.3% | 90%* | $1.2 \times 10^6$ | 4%* |
| Chen17 [89] | PP fiber[†] | 98.9% | – | 18 200 | 2.5%* |
| **Spectrally engineered** | | | | | |
| Meyer-Scott13 [56] | Crossed PM fiber | 92.2% | 70%* | 1300* | 20% |
| **Quantum dots** | | | | | |
| Huber17 [28] | GaAs quantum dot | 94% | 67% | – | ≪ 1% |
| Jöns17 [29] | InAsP quantum dot | 81.7% | – | – | 0.1%* |
| **This work** **Chip 1** | PPKTP WG + Sagnac[†] | 95.8% | 82% | $3.5 \times 10^6$ | 43% |
| **Chip 2** | PPKTP WG + Sagnac[†] | 98.8% | 89% | $5.6 \times 10^6$ | 27% |

### Funding

### Acknowledgments

We would like to thank Johannes Tiedau, Geoff Pryde, Morgan Weston, Deny Hamel, Thomas Jennewein, and Sergei Slussarenko for helpful discussions.

### References

1. M. Giustina, M. A. M. Versteegh, S. Wengerowsky, J. Handsteiner, A. Hochrainer, K. Phelan, F. Steinlechner, J. Kofler, J.-A. Larsson, C. Abellán, W. Amaya, V. Pruneri, M. W. Mitchell, J. Beyer, T. Gerrits, A. E. Lita, L. K. Shalm, S. W. Nam, T. Scheidl, R. Ursin, B. Wittmann, and A. Zeilinger, "Significant-loophole-free test of Bell's theorem with entangled photons," Phys. Rev. Lett. **115**, 250401 (2015).
2. L. K. Shalm, E. Meyer-Scott, B. G. Christensen, P. Bierhorst, M. A. Wayne, M. J. Stevens, T. Gerrits, S. Glancy, D. R. Hamel, M. S. Allman, K. J. Coakley, S. D. Dyer, C. Hodge, A. E. Lita, V. B. Verma, C. Lambrocco, E. Tortorici, A. L. Migdall, Y. Zhang, D. R. Kumor, W. H. Farr, F. Marsili, M. D. Shaw, J. A. Stern, C. Abellán, W. Amaya, V. Pruneri, T. Jennewein, M. W. Mitchell, P. G. Kwiat, J. C. Bienfang, R. P. Mirin, E. Knill, and S. W. Nam, "Strong loophole-free test of local realism," Phys. Rev. Lett. **115**, 250402 (2015).
3. B. G. Christensen, Y.-C. Liang, N. Brunner, N. Gisin, and P. G. Kwiat, "Exploring the limits of quantum nonlocality with entangled photons," Phys. Rev. X **5**, 041052 (2015).
4. H. S. Poh, S. K. Joshi, A. Cerè, A. Cabello, and C. Kurtsiefer, "Approaching Tsirelson's bound in a photon pair experiment," Phys. Rev. Lett. **115**, 180408 (2015).
5. R. Ursin, F. Tiefenbacher, T. Schmitt-Manderbach, H. Weier, T. Scheidl, M. Lindenthal, B. Blauensteiner, T. Jennewein, J. Perdigues, P. Trojek, B. Omer, M. Furst, M. Meyenburg, J. Rarity, Z. Sodnik, C. Barbieri, H. Weinfurter, and A. Zeilinger, "Entanglement-based quantum communication over 144 km," Nat. Phys. **3**, 481–486 (2007).
6. T. Scheidl, R. Ursin, A. Fedrizzi, S. Ramelow, X.-S. Ma, T. Herbst, R. Prevedel, L. Ratschbacher, J. Kofler, T. Jennewein, and A. Zeilinger, "Feasibility of 300 km quantum key distribution with entangled states," New J. Phys. **11**, 085002 (2009).
7. J. Yin, Y. Cao, Y.-H. Li, S.-K. Liao, L. Zhang, J.-G. Ren, W.-Q. Cai, W.-Y. Liu, B. Li, H. Dai, G.-B. Li, Q.-M. Lu, Y.-H. Gong, Y. Xu, S.-L. Li, F.-Z. Li, Y.-Y. Yin, Z.-Q. Jiang, M. Li, J.-J. Jia, G. Ren, D. He, Y.-L. Zhou, X.-X. Zhang, N. Wang, X. Chang, Z.-C. Zhu, N.-L. Liu, Y.-A. Chen, C.-Y. Lu, R. Shu, C.-Z. Peng, J.-Y. Wang, and J.-W. Pan, "Satellite-based entanglement distribution over 1200 kilometers," Science **356**, 1140–1144 (2017).
8. T. Nagata, R. Okamoto, J. L. O'Brien, K. Sasaki, and S. Takeuchi, "Beating the standard quantum limit with four-entangled photons," Science **316**, 726–729 (2007).
9. S. Slussarenko, M. M. Weston, H. M. Chrzanowski, L. K. Shalm, V. B. Verma, S. W. Nam, and G. J. Pryde, "Unconditional violation of the shot-noise limit in photonic quantum metrology," Nat. Photonics **11**, 700–703 (2017).
10. K. Wang, X. Wang, X. Zhan, Z. Bian, J. Li, B. C. Sanders, and P. Xue, "Entanglement-enhanced quantum metrology in a noisy environment," Phys. Rev. A **97**, 042112 (2018).
11. R. S. Bennink, "Optimal collinear Gaussian beams for spontaneous parametric down-conversion," Phys. Rev. A **81**, 053805 (2010).
12. M. Tillmann, B. Dakic, R. Heilmann, S. Nolte, A. Szameit, and P. Walther, "Experimental boson sampling," Nat Photon **7**, 540–544 (2013).
13. A. Crespi, R. Osellame, R. Ramponi, D. J. Brod, E. F. Galvao, N. Spagnolo, C. Vitelli, E. Maiorino, P. Mataloni, and F. Sciarrino, "Integrated multimode interferometers with arbitrary designs for photonic boson sampling," Nat Photon **7**, 545–549 (2013).
14. J. Carolan, C. Harrold, C. Sparrow, E. Martín-López, N. J. Russell, J. W. Silverstone, P. J. Shadbolt, N. Matsuda, M. Oguma, M. Itoh, G. D. Marshall, M. G. Thompson, J. C. F. Matthews, T. Hashimoto, J. L. O'Brien, and A. Laing, "Universal linear optics," Science **349**, 711–716 (2015).
15. R. Valivarthi, M. l. G. Puigibert, Q. Zhou, G. H. Aguilar, V. B. Verma, F. Marsili, M. D. Shaw, S. W. Nam, D. Oblak, and W. Tittel, "Quantum teleportation across a metropolitan fibre network," Nat Photon **10**, 676–680 (2016).
16. Q.-C. Sun, Y.-L. Mao, S.-J. Chen, W. Zhang, Y.-F. Jiang, Y.-B. Zhang, W.-J. Zhang, S. Miki, T. Yamashita, H. Terai, X. Jiang, T.-Y. Chen, L.-X. You, X.-F. Chen, Z. Wang, J.-Y. Fan, Q. Zhang, and J.-W. Pan, "Quantum teleportation with independent sources and prior entanglement distribution over a network," Nat Photon **10**, 671–675 (2016).
17. X.-L. Wang, L.-K. Chen, W. Li, H.-L. Huang, C. Liu, C. Chen, Y.-H. Luo, Z.-E. Su, D. Wu, Z.-D. Li, H. Lu, Y. Hu, X. Jiang, C.-Z. Peng, L. Li, N.-L. Liu, Y.-A. Chen, C.-Y. Lu, and J.-W. Pan, "Experimental ten-photon entanglement," Phys. Rev. Lett. **117**, 210502 (2016).
18. L.-K. Chen, Z.-D. Li, X.-C. Yao, M. Huang, W. Li, H. Lu, X. Yuan, Y.-B. Zhang, X. Jiang, C.-Z. Peng, L. Li, N.-L. Liu, X. Ma, C.-Y. Lu, Y.-A. Chen, and J.-W. Pan, "Observation of ten-photon entanglement using thin $BiB_3O_6$ crystals," Optica **4**, 77–83 (2017).

19. H. Herrmann, X. Yang, A. Thomas, A. Poppe, W. Sohler, and C. Silberhorn, "Post-selection free, integrated optical source of non-degenerate, polarization entangled photon pairs," Opt. Express **21**, 27981–27991 (2013).

20. L. Sansoni, K. H. Luo, C. Eigner, R. Ricken, V. Quiring, H. Herrmann, and C. Silberhorn, "A two-channel, spectrally degenerate polarization entangled source on chip," npj Quantum Inf. **3**, 5 (2017).

21. P. Vergyris, F. Kaiser, E. Gouzien, G. Sauder, T. Lunghi, and S. Tanzilli, "Fully guided-wave photon pair source for quantum applications," Quantum Sci. Technol. **2**, 024007 (2017).

22. X. Li, P. L. Voss, J. E. Sharping, and P. Kumar, "Optical-fiber source of polarization-entangled photons in the 1550 nm telecom band," Phys. Rev. Lett. **94**, 053601 (2005).

23. J. Fan, M. D. Eisaman, and A. Migdall, "Bright phase-stable broadband fiber-based source of polarization-entangled photon pairs," Phys. Rev. A **76**, 043836 (2007).

24. H. Takesue, H. Fukuda, T. Tsuchizawa, T. Watanabe, K. Yamada, Y. Tokura, and S. Itabashi, "Generation of polarization entangled photon pairs using silicon wirewaveguide," Opt. Express **16**, 5721–5727 (2008).

25. Y.-H. Li, Z.-Y. Zhou, L.-T. Feng, W.-T. Fang, S.-l. Liu, S.-K. Liu, K. Wang, X.-F. Ren, D.-S. Ding, L.-X. Xu, and B.-S. Shi, "On-chip multiplexed multiple entanglement sources in a single silicon nanowire," Phys. Rev. Appl. **7**, 064005 (2017).

26. M. Zhang, L.-T. Feng, Z.-Y. Zhou, Y. Chen, H. Wu, M. Li, G.-P. Guo, G.-C. Guo, D.-X. Dai, and X.-F. Ren, "Generation of multiphoton entangled quantum states with a single silicon nanowire," ArXiv:1803.01641 (2018).

27. W.-T. Fang, Y.-H. Li, Z.-Y. Zhou, L.-X. Xu, G.-C. Guo, and B.-S. Shi, "On-chip generation of time-and wavelength-division multiplexed multiple time-bin entanglement," Opt. Express **26**, 12912–12921 (2018).

28. D. Huber, M. Reindl, Y. Huo, H. Huang, J. S. Wildmann, O. G. Schmidt, A. Rastelli, and R. Trotta, "Highly indistinguishable and strongly entangled photons from symmetric GaAs quantum dots," Nat. Commun. **8**, 15506 (2017).

29. K. D. Jöns, L. Schweickert, M. A. M. Versteegh, D. Dalacu, P. J. Poole, A. Gulinatti, A. Giudice, V. Zwiller, and M. E. Reimer, "Bright nanoscale source of deterministic entangled photon pairs violating Bell's inequality," Sci. Rep. **7**, 1700 (2017).

30. B. G. Christensen, K. T. McCusker, J. B. Altepeter, B. Calkins, T. Gerrits, A. E. Lita, A. Miller, L. K. Shalm, Y. Zhang, S. W. Nam, N. Brunner, C. C. W. Lim, N. Gisin, and P. G. Kwiat, "Detection-loophole-free test of quantum nonlocality, and applications," Phys. Rev. Lett. **111**, 130406 (2013).

31. D. N. Klyshko, "Use of two-photon light for absolute calibration of photoelectric detectors," Sov. J. Quantum Electron. **10**, 1112 (1980).

32. S. Ramelow, A. Mech, M. Giustina, S. Gröblacher, W. Wieczorek, J. Beyer, A. Lita, B. Calkins, T. Gerrits, S. W. Nam, A. Zeilinger, and R. Ursin, "Highly efficient heralding of entangled single photons," Opt. Express **21**, 6707–6717 (2013).

33. M. Giustina, A. Mech, S. Ramelow, B. Wittmann, J. Kofler, J. Beyer, A. Lita, B. Calkins, T. Gerrits, S. W. Nam, R. Ursin, and A. Zeilinger, "Bell violation using entangled photons without the fair-sampling assumption," Nature **497**, 227–230 (2013).

34. M. M. Weston, H. M. Chrzanowski, S. Wollmann, A. Boston, J. Ho, L. K. Shalm, V. B. Verma, M. S. Allman, S. W. Nam, R. B. Patel, S. Slussarenko, and G. J. Pryde, "Efficient and pure femtosecond-pulse-length source of polarization-entangled photons," Opt. Express **24**, 10869–10879 (2016).

35. P. B. Dixon, D. Rosenberg, V. Stelmakh, M. E. Grein, R. S. Bennink, E. A. Dauler, A. J. Kerman, R. J. Molnar, and F. N. C. Wong, "Heralding efficiency and correlated-mode coupling of near-IR fiber-coupled photon pairs," Phys. Rev. A **90**, 043804 (2014).

36. W. P. Grice, R. S. Bennink, D. S. Goodman, and A. T. Ryan, "Spatial entanglement and optimal single-mode coupling," Phys. Rev. A **83**, 023810 (2011).

37. T. Guerreiro, A. Martin, B. Sanguinetti, N. Bruno, H. Zbinden, and R. T. Thew, "High efficiency coupling of photon pairs in practice," Opt. Express **21**, 27641–27651 (2013).

38. L. G. Helt, M. Liscidini, and J. E. Sipe, "How does it scale? comparing quantum and classical nonlinear optical processes in integrated devices," J. Opt. Soc. Am. B **29**, 2199–2212 (2012).

39. A. Christ, K. Laiho, A. Eckstein, T. Lauckner, P. J. Mosley, and C. Silberhorn, "Spatial modes in waveguided parametric down-conversion," Phys. Rev. A **80**, 033829 (2009).

40. A. Eckstein, A. Christ, P. J. Mosley, and C. Silberhorn, "Highly efficient single-pass source of pulsed single-mode twin beams of light," Phys. Rev. Lett. **106**, 013603 (2011).

41. D. Ljunggren and M. Tengner, "Optimal focusing for maximal collection of entangled narrow-band photon pairs into single-mode fibers," Phys. Rev. A **72**, 062301 (2005).

42. M. Fiorentino, S. M. Spillane, R. G. Beausoleil, T. D. Roberts, P. Battle, and M. W. Munro, "Spontaneous parametric down-conversion in periodically poled KTP waveguides and bulk crystals," Opt. Express **15**, 7479–7488 (2007).

43. N. Montaut, L. Sansoni, E. Meyer-Scott, R. Ricken, V. Quiring, H. Herrmann, and C. Silberhorn, "High-efficiency plug-and-play source of heralded single photons," Phys. Rev. Appl. **8**, 024021 (2017).

44. Q.-C. Sun, Y.-L. Mao, Y.-F. Jiang, Q. Zhao, S.-J. Chen, W. Zhang, W.-J. Zhang, X. Jiang, T.-Y. Chen, L.-X. You, L. Li, Y.-D. Huang, X.-F. Chen, Z. Wang, X. Ma, Q. Zhang, and J.-W. Pan, "Entanglement swapping with independent sources over an optical-fiber network," Phys. Rev. A **95**, 032306 (2017).

45. Q. Wang and X.-B. Wang, "Efficient implementation of the decoy-state measurement-device-independent quantum key distribution with heralded single-photon sources," Phys. Rev. A **88**, 052332 (2013).

46. J. Flórez, O. Calderón, A. Valencia, and C. I. Osorio, "Correlation control for pure and efficiently generated heralded single photons," Phys. Rev. A **91**, 013819 (2015).

47. E. Meyer-Scott, N. Montaut, J. Tiedau, L. Sansoni, H. Herrmann, T. J. Bartley, and C. Silberhorn, "Limits on the heralding efficiencies and spectral purities of spectrally filtered single photons from photon-pair sources," Phys. Rev. A **95**, 061803(R) (2017).

48. W. P. Grice and I. A. Walmsley, "Spectral information and distinguishability in type-II down-conversion with a broadband pump," Phys. Rev. A **56**, 1627–1634 (1997).

49. P. J. Mosley, J. S. Lundeen, B. J. Smith, P. Wasylczyk, A. B. U'Ren, C. Silberhorn, and I. A. Walmsley, "Heralded generation of ultrafast single photons in pure quantum states," Phys. Rev. Lett. **100**, 133601 (2008).

50. N. Bruno, A. Martin, T. Guerreiro, B. Sanguinetti, and R. T. Thew, "Pulsed source of spectrally uncorrelated and indistinguishable photons at telecom wavelengths," Opt. Express **22**, 17246–17253 (2014).

51. S. Arahira, N. Namekata, T. Kishimoto, H. Yaegashi, and S. Inoue, "Generation of polarization entangled photon pairs at telecommunication wavelength using cascaded $\chi^{(2)}$ processes in a periodically poled LiNbO$_3$ ridge waveguide," Opt. Express **19**, 16032–16043 (2011).

52. C. Clausen, F. Bussières, A. Tiranov, H. Herrmann, C. Silberhorn, W. Sohler, M. Afzelius, and N. Gisin, "A source of polarization-entangled photon pairs interfacing quantum memories with telecom photons," New J. Phys. **16**, 093058 (2014).

53. C. Söller, O. Cohen, B. J. Smith, I. A. Walmsley, and C. Silberhorn, "High-performance single-photon generation with commercial-grade optical fiber," Phys. Rev. A **83**, 031806 (2011).

54. R. J. A. Francis-Jones, R. A. Hoggarth, and P. J. Mosley, "All-fiber multiplexed source of high-purity single photons," Optica **3**, 1270–1273 (2016).

55. Y. Tsujimoto, M. Tanaka, N. Iwasaki, R. Ikuta, S. Miki, T. Yamashita, H. Terai, T. Yamamoto, M. Koashi, and N. Imoto, "High-fidelity entanglement swapping and generation of three-qubit GHZ state using asynchronous telecom photon pair sources," Sci. Reports **8**, 1446 (2018).

56. E. Meyer-Scott, V. Roy, J.-P. Bourgoin, B. L. Higgins, L. K. Shalm, and T. Jennewein, "Generating polarization-entangled photon pairs using cross-spliced birefringent fibers," Opt. Express **21**, 6205–6212 (2013).

57. O. Kuzucu, F. N. C. Wong, S. Kurimura, and S. Tovstonog, "Joint temporal density measurements for two-photon state characterization," Phys. Rev. Lett. **101**, 153602 (2008).

58. B.-S. Shi and A. Tomita, "Generation of a pulsed polarization entangled photon pair using a Sagnac interferometer," Phys. Rev. A **69**, 013803 (2004).

59. T. Kim, M. Fiorentino, and F. N. C. Wong, "Phase-stable source of polarization-entangled photons using a polarization Sagnac interferometer," Phys. Rev. A **73**, 012316 (2006).

60. D. Hamel, "Realization of novel entangled photon sources using periodically poled materials," Master's thesis, University of Waterloo (2010).

61. G. Brida, M. Chekhova, M. Genovese, and L. Krivitsky, "Generation of different Bell states within the spontaneous parametric down-conversion phase-matching bandwidth," Phys. Rev. A **76**, 053807 (2007).

62. B. Fang, M. Liscidini, J. E. Sipe, and V. O. Lorenz, "Multidimensional characterization of an entangled photon-pair source via stimulated emission tomography," Opt. Express **24**, 10013–10019 (2016).

63. M. Avenhaus, A. Eckstein, P. J. Mosley, and C. Silberhorn, "Fiber-assisted single-photon spectrograph," Opt. Lett. **34**, 2873–2875 (2009).

64. T. Gerrits, M. J. Stevens, B. Baek, B. Calkins, A. Lita, S. Glancy, E. Knill, S. W. Nam, R. P. Mirin, R. H. Hadfield, R. S. Bennink, W. P. Grice, S. Dorenbos, T. Zijlstra, T. Klapwijk, and V. Zwiller, "Generation of degenerate, factorizable, pulsed squeezed light at telecom wavelengths," Opt. Express **19**, 24434–24447 (2011).

65. C. K. Hong, Z. Y. Ou, and L. Mandel, "Measurement of subpicosecond time intervals between two photons by interference," Phys. Rev. Lett. **59**, 2044–2046 (1987).

66. F. Kaneda, K. Garay-Palmett, A. B. U'Ren, and P. G. Kwiat, "Heralded single-photon source utilizing highly nondegenerate, spectrally factorable spontaneous parametric downconversion," Opt. Express **24**, 10733–10747 (2016).

67. C. I. Osorio, N. Sangouard, and R. T. Thew, "On the purity and indistinguishability of down-converted photons," J. Phys. B: At. Mol. Opt. Phys. **46**, 055501 (2013).

68. J. R. Johansson, P. D. Nation, and F. Nori, "QuTiP: An open-source Python framework for the dynamics of open quantum systems," Comput. Phys. Commun. **183**, 1760–1772 (2012).

69. D. F. V. James, P. G. Kwiat, W. J. Munro, and A. G. White, "Measurement of qubits," Phys. Rev. A **64**, 052312 (2001).

70. S. M. Jobling, "Adaptive optics for improved mode-coupling efficiencies," Master's thesis, University of Illinois at Urbana-Champaign (2008).

71. A. M. Brańczyk, A. Fedrizzi, T. M. Stace, T. C. Ralph, and A. G. White, "Engineered optical nonlinearity for quantum light sources," Opt. Express **19**, 55–65 (2011).

72. P. B. Dixon, J. H. Shapiro, and F. N. C. Wong, "Spectral engineering by Gaussian phase-matching for quantum photonics," Opt. Express **21**, 5879–5890 (2013).

73. F. Graffitti, P. Barrow, M. Proietti, D. Kundys, and A. Fedrizzi, "Independent high-purity photons created in domain-engineered crystals," Optica **5**, 514–517 (2018).

74. M. Santandrea, C. Silberhorn, et al. are preparing a manuscript to be titled, "Fabrication limits of waveguides in

nonlinear crystals and their impact on quantum optics applications," (2018).

75. D. G. Matei, T. Legero, S. Häfner, C. Grebing, R. Weyrich, W. Zhang, L. Sonderhouse, J. M. Robinson, J. Ye, F. Riehle, and U. Sterr, "1.5 $\mu m$ lasers with sub-10 mHz linewidth," Phys. Rev. Lett. **118**, 263202 (2017).

76. N. Korolkova, G. Leuchs, R. Loudon, T. C. Ralph, and C. Silberhorn, "Polarization squeezing and continuous-variable polarization entanglement," Phys. Rev. A **65**, 052306 (2002).

77. T. Iskhakov, M. V. Chekhova, and G. Leuchs, "Generation and direct detection of broadband mesoscopic polarization-squeezed vacuum," Phys. Rev. Lett. **102**, 183602 (2009).

78. T. S. Iskhakov, I. N. Agafonov, M. V. Chekhova, and G. Leuchs, "Polarization-entangled light pulses of $10^5$ photons," Phys. Rev. Lett. **109**, 150502 (2012).

79. P. G. Kwiat, K. Mattle, H. Weinfurter, A. Zeilinger, A. V. Sergienko, and Y. Shih, "New high-intensity source of polarization-entangled photon pairs," Phys. Rev. Lett. **75**, 4337–4341 (1995).

80. P. G. Kwiat, E. Waks, A. G. White, I. Appelbaum, and P. H. Eberhard, "Ultrabright source of polarization-entangled photons," Phys. Rev. A **60**, R773–R776 (1999).

81. M. Fiorentino, G. Messin, C. E. Kuklewicz, F. N. C. Wong, and J. H. Shapiro, "Generation of ultrabright tunable polarization entanglement without spatial, spectral, or temporal constraints," Phys. Rev. A **69**, 041801 (2004).

82. F. Steinlechner, M. Gilaberte, M. Jofre, T. Scheidl, J. P. Torres, V. Pruneri, and R. Ursin, "Efficient heralding of polarization-entangled photons from type-0 and type-II spontaneous parametric downconversion in periodically poled KTiOPO$_4$," J. Opt. Soc. Am. B **31**, 2068–2076 (2014).

83. Y. Chen, S. Ecker, S. Wengerowsky, L. Bulla, S. Koduru Joshi, F. Steinlechner, and R. Ursin, "Polarization entanglement by time-reversed Hong-Ou-Mandel interference," arXiv: 1807.00383 (2018).

84. P. G. Evans, R. S. Bennink, W. P. Grice, T. S. Humble, and J. Schaake, "Bright source of spectrally uncorrelated polarization-entangled photons with nearly single-mode emission," Phys. Rev. Lett. **105**, 253601 (2010).

85. R.-B. Jin, R. Shimizu, K. Wakui, M. Fujiwara, T. Yamashita, S. Miki, H. Terai, Z. Wang, and M. Sasaki, "Pulsed Sagnac polarization-entangled photon source with a PPKTP crystal at telecom wavelength," Opt. Express **22**, 11498–11507 (2014).

86. C. Autebert, J. Trapateau, A. Orieux, A. Lemaître, C. Gomez-Carbonell, E. Diamanti, I. Zaquine, and S. Ducci, "Multi-user quantum key distribution with entangled photons from an AlGaAs chip," Quantum Sci. Technol. **1**, 01LT02 (2016).

87. S. Atzeni, A. S. Rab, G. Corrielli, E. Polino, M. Valeri, P. Mataloni, N. Spagnolo, A. Crespi, F. Sciarrino, and R. Osellame, "Integrated sources of entangled photons at the telecom wavelength in femtosecond-laser-written circuits," Optica **5**, 311–314 (2018).

88. H. C. Lim, A. Yoshizawa, H. Tsuchida, and K. Kikuchi, "Stable source of high quality telecom-band polarization-entangled photon-pairs based on a single, pulse-pumped, short PPLN waveguide," Opt. Express **16**, 12460–12468 (2008).

89. C. Chen, E. Y. Zhu, A. Riazi, A. V. Gladyshev, C. Corbari, M. Ibsen, P. G. Kazansky, and L. Qian, "Compensation-free broadband entangled photon pair sources," Opt. Express **25**, 22667–22678 (2017).